\documentclass[12pt]{article}
\usepackage{pic04}
\usepackage{hyperref}%%[backref]
\usepackage{url}
\usepackage{graphicx}
\usepackage{epsfig}

\def\ie{{\it\kern-2pt i.\kern-.5pt e.\kern-2pt}}  
  
\def\up#1{$^{#1}$}  
\def\ifm#1{\relax\ifmmode#1\else$#1$\fi}
\def\Bbar{\ifm{\rlap{\kern.22em\raise1.9ex\hbox to.58em{\hrulefill}} B}}
   
 \def\bra#1,{\ifm{\langle\,#1\,|}} %%\def\hb{\vphantom B} 

\def\to{\ifm{\rightarrow}} \def\sig{\ifm{\sigma}}   \def\plm{\ifm{\pm}}
  
\def\ff{$\phi$--factory}    \def\f{\ifm{\phi}} 
 \def\pic{\ifm{\pi^+\pi^-}} \def\pio{\ifm{\pi^0\pi^0}} 
  
\def\po{\ifm{\pi^0}}
\def\ks{\ifm{K_S}} \def\kl{\ifm{K_L}} 
  
\def\eps{\ifm{\epsilon}} \def\epm{\ifm{e^+e^-}}
    
\def\Kb{\ifm{\rlap{\kern.3em\raise1.9ex\hbox to.6em{\hrulefill}} K}}

\def\noc{\relax\hglue0pt{\rlap{$C$}\raise.15ex\hbox{$\kern
.18em\backslash$}}}
\def\nop{\relax\hglue0pt{\rlap{$P$}\raise.15ex\hbox{$\kern
.18em\backslash$}}}
\def\noT{\relax\hglue0pt{\rlap{$T$}\raise.15ex\hbox{$\kern
.18em\backslash$}}}
  
\def\ko{\ifm{K^0}}  
\def\gam{\ifm{\gamma}}  
 \def\ab{\ifm{\sim}}  \def\x{\ifm{\times}}
\def\sta#1,{\ifm{|\,#1\,\rangle}} \def\ket#1,{\ifm{|\,#1\,\rangle}} 
  
\def\pt#1,#2,{\ifm{#1\x10^{#2}}}
  
\def\minus{$-$}  \def\dif{\hbox{d}}   
\def\bye{\end{document}}
\font\euler=eufm10 at 12pt
\def\Ma{\hbox{\euler M}}
\def\eiii{\ifm{\pi^\pm e^\mp\nu}}  
\def\muiii{\ifm{\pi^\pm \mu^\mp\nu}}

\catcode`@=11 %%At signs are letters
\newskip\z@skip \z@skip=0pt plus0pt minus0pt
\def\m@th{\mathsurround=\z@}
\def\ialign{\everycr{}\tabskip\z@skip\halign} % initialized \halign
\def\eqalign#1{\null\,\vcenter{\openup\jot\m@th
  \ialign{\strut\hfil$\displaystyle{##}$&$\displaystyle{{}##}$\hfil
      \crcr#1\crcr}}\,}
\catcode`@=12 % at signs are no longer letters
\let\cl=\centerline

\def\figbox#1;#2;{\parbox{#2cm}{\epsfig{file=#1.eps,width=#2cm}}}
\def\figboxc#1;#2;{\cl{\figbox#1;#2;}}

\begin{document}

\title{\bf Kaon Decays and $V_{us}$}
\author{Paolo Franzini\\[3mm]
Dipartimento di Fisica, Universita' di Roma, {\em La Sapienza}\\%%
P.le A. Moro, Roma Italy\\
e-mail:paolo.franzini@lnf.infn.it}
\maketitle

\baselineskip=14.5pt
\begin{abstract}
Unitarity for the first row of the quark mixing matrix appears to be fully satisfied by the value of $V_{us}$ obtained from recent new measurements of kaon semileptonic decay rates.
\end{abstract}
\thispagestyle{empty}
\newpage

\baselineskip=17pt

\section{Introduction}\noindent
%%\citen{cab} 
What is today called quark mixing was first introduced in 1963 by Cabibbo \cite{ref:cab} as an angle characterizing the relative strength of strangeness changing and strangeness conserving weak amplitudes. In modern language we write the quark weak current as $J_\alpha=\overline{\vphantom{|}\mathbf U}\gam_\alpha(1-\gam_5){\mathbf V}{\mathbf D}$, where ${\bf U, D}$ are the charge 2/3, \minus1/3 quark vectors and ${\bf V}$ is a unitary matrix \cite{ref:km}, often called the CKM matrix. Whether ${\bf V}$ is unitary can be checked experimentally.

For no good reason at all it became fashionable, in the mid-late nineties, to expect the CKM-matrix not to be unitary, providing great impetus to the study of $B$ mesons. The beautiful experiments at KEK and SLAC have dashed such expectation, verifying unitarity to ${\cal O}(10\%)$. By far the best check comes however from kaon physics.
The so-called first row unitarity requires $|V_{ud}|^2+|V_{us}|^2+|V_{ub}|^2=1$ which, because $|V_{ub}|^2\ab10^{-5}$ \cite{ref:vub}, is equivalent to $|V_{ud}|^2+|V_{us}|^2=1$.
Semileptonic kaon decays offer possibly the cleanest way to obtain an accurate value of the Cabibbo angle or better $|V_{us}|$. Since $K\to\pi$ is a 0\up-\to0\up- transition, only the vector part of the weak current has a non vanishing contribution. Such processes are protected by the Ademollo-Gatto theorem against SU(3) breaking corrections to lowest order in $m_s-m_d$. Recent advances in lattice gauge techniques have allowed the computing of the pseudoscalar decay constants $f_\pi$ and $f_\mu$. As a consequence, the $K\to\mu\nu$ partial decay rate provides a new handle for the determination of $|V_{us}|$.%

\section{Experimental situation prior to 2004}
For several years the Particle Data Group (PDG) has reported values of $|V_{ud}|$ and $|V_{us}|$ in slight disagreement with unitarity. The 2004 version \cite{ref:pdg}, gives $|V_{us}|$=0.2200\break\plm0.0026 and $|V_{ud}|$=0.9738\plm0.0005 from which the sum of the squares is 0.9966\break\plm0.0015 which verifies unitarity to ${\cal O}(0.33\%)$ albeit with a deviation of \ab2\sig. This result is however based on the so called PDG fit. Examining inputs used for the fit and the output of the fit, one notices a few peculiarities:
\begin{enumerate}
\item Direct measurements of the partial rates and/or branching ratios for the relevant decays, $K\to\pi e\nu,\pi\mu\nu$, are quite old, with accuracies of ${\cal O}(5\%)$
\item By using ratios of partial rates, even far removed from the quantities of interest, the fit comes out with errors as low as 0.7\%, this being the case for \kl\to$\pi^\pm e^\mp\nu$
\item Radiative contribution to the input values are essentially unknown. This invalidates the constraints used in the fit as well as the usefulness of the resulting BR values for obtaining $|V_{us}|$
\item There are large differences in the form factor slopes for neutral and charged kaons. The ratio $K_{e3}/K_{\mu3}$ which is a unique function of the form factor slope parameters is quite inconsistent between charge and neutral kaons. The most recent results, from KEK \cite{ref:kek} and Istra+ \cite{ref:istrae,ref:istramu}, clearly indicate a smaller ratio.
\item The charged kaon lifetime, given by PDG with an accuracy of 0.2\% is in fact an average or fit of values covering an interval of 1.5\%. While PDG might wish to be democratic, some of those values must be wrong.
\end{enumerate}
Both central values and errors given by PDG for branching ratios are therefore suspect and anyway not usable, because of ignorance about inclusion of radiation, to obtain $|V_{us}|$ to better than several per cent.
\section{How to get to $|V_{us}|$}
In principle the connection between partial decay rates and $|V_{us}|$ is as simple as:
\begin{equation}
\Gamma(K\to\pi\ell_{\rm i}\nu)={\cal N}\;|V_{us}|^2\int \overline{|\Ma|}^{\:2}\dif E_1\dif E_2\x{\cal C}
\label{gamma}
\end{equation}
where ${\cal N}$ contains well known things like the Fermi constants, some $\pi$'s and two's, Clebsches etc. There is a phase space integral and some corrections ${\cal C}$. We begin with things that follow from measurement. Partial width are always obtained from $\Gamma={\rm BR}/\tau$,\, \ie\ \, we must measure both branching fractions and lifetimes. The phase space integral is a function of masses, quite well known and of the shape of the form factor.
The $f_+$ form factor is defined by: $\langle\pi|J^V_\alpha|K\rangle\equiv f_+(t=0)\x\tilde f_+(t)(P+p)_\alpha$ where $\tilde f(t)_+=1+\lambda'_+t/m+(\lambda''_+/2)(t/m)^2+..$ or some other parameterization, with $\tilde f_+(0)=1$. $P,p$ are the initial and final four momenta, $m$ is the pion mass and $J_\alpha^V$ is the vector part of the quark current. The form factor parameters can and must be measured for the vector form factor $f_+$ and also for the scalar form factor $f_0$. In the following we retain the three parameters: $\lambda'_+,\lambda''_+$ defined above and the analogous parameter $\lambda_0.$

Amongst the corrections, radiative effects are quite important. Since we cannot turn off electromagnetism, we must include in the measurement some well defined fraction of the radiative decays. Ideally we want to measure the rate or BR for decays to $\pi e\nu$ {\it etc}., plus any number of \gam's. This rate is finite and calculable. A very important point, being only recently fully recognized, is that radiative effects must be included in the acceptance calculations, in practice in the decay generators used for the Monte Carlo simulations. Otherwise counting mistakes of several per cent are inevitable.

There is more to the corrections. There are radiative corrections, both short distance and low energy corrections to the phase space density both with and without external radiation. There are isospin breaking correction to account for \po-$\pi^\pm$ and $K^0$-$K^\pm$ (or $u$-$d$) differences. Finally pions and kaons are not quite the same objects and $SU(3)_F$ breaking (or $s$-$d$ difference induced) corrections must be included. This last corrections is the factor $f_+(0)$, mentioned above in the definition of the form factor. We can note that there is a unique value $f^{K^0}_+(0)$ for both $f_+$ and $f_0$ in neutral kaon decays and another for charged kaons.
One ends up with something like:\footnote{The numerical factor 768=2\up8\x3, also appears sometimes as 128, other times as 192. They are all equally valid and the difference is made up by the phase space integrals $I_i$.}
\begin{equation}
\Gamma={G_{\rm F}^2\:M_K^5\over768\:\pi^3}\:|V_{us}|^2\:S_{\rm EW}\:|f^{\ko}_+(0)|^2\:C_K^2\:I^\ell_K \:[1+\delta^K_{SU(2)}+\delta^K_{em}]
\label{gamma1}
\end{equation}
$I_{i}(\lambda', \lambda'', \lambda_0)$ is the integral of the phase space density, factoring out $f_+^{K^0}$ and without radiative corrections.
Short distance radiative corrections are in the Sirlin term $S_{\rm EW}$ \cite{ref:asir}. In addition long distance radiative corrections \cite{ref:cir1,ref:cir2} for form factor, phase space density and $SU(2)$ breaking are included as $\delta^{i}_{\rm em}$ and $\Delta I_{i} (\lambda)$.
\section{New indications}\label{sec:new}
The first evidence that not all was correct with the PDG information came from E865 at BNL, who published in 2003 a fully inclusive value for $K^+_{e3}$ decays \cite{ref:e865} of BR($K^+\to\po e^+\nu(\gam)$)=(5.13\plm0.10)\%, to be compared to (4.87\plm0.06)\%. The difference is not very large and moreover E865 actually measures the ratio $e3/(\pi2+\mu3+\pi3)$ and relies on PDG for knowledge of the denominator. 
Early in 2004 KLOE \cite{ref:tom} reported on $\Gamma(\ks\to\pi^\pm e^\mp\nu(\gam))$, finding $\Gamma$=\pt(7.92\pm0.12),6, s\up{-1}, to be compared to \pt(7.50\pm0.08),6, s\up{-1} for \kl. This result does not rely in any way on external information but uses the new, precise KLOE result for $\Gamma(\pic)/\Gamma(\pio)$ \cite{ref:pipo}. The $K^+$ results implies $|V_{us}|$=0.2272\plm0.0029 while from KLOE $|V_{us}|$=0.2254\plm0.0026, both values in very good agreement with unitarity which requires a value of 0.2265\plm0.0022.
\section{Doing it correctly}
Past experience proves the dangers of using too many intermediate steps and information from different experiments. The best way to obtain partial rate measurements is to start with a tagged kaon sample and count decays to the appropriate channels. Accurate lifetimes are still lacking, especially for \kl-mesons but also possibly for charged kaons. Finally, to extract $|V_{us}|$ the form factor parameters must be measured, values calculated in chiral perturbation not being reliable.

An alternate way consists of measuring all major ratios $\Gamma_i/\tilde\Gamma$, where $\Gamma_i$ is the partial width for channel $i$ and $\tilde\Gamma$ is the partial width for a major channel, not included in the previous set. Then the constraint $\tilde\Gamma+\sum\Gamma_i=1-\eps$ allows solving for all branching ratios $\Gamma_i/\Gamma$ and $\tilde\Gamma/\Gamma$. For the \kl\ case, choosing the 4 channels $\pi e\nu$, $\pi\mu\nu$, \pic\po\ and \pio\po\ we have 0.0035$\,<\!\eps\!<\,$0.0036 \ie\ one can reach in principle an accuracy of \ab0.0001 for $K_{e3}$ and $K_{\mu3}$ or \ab0.03\%.

The first method is ideally employed at a \ff\ where the production of the \ks\kl\ pair in \f-decay provides a tagged, monochromatic \kl-beam of known flux. This is the choice of KLOE \cite{ref:kl04} which has obtained preliminary results discussed below. An advantage of working with a tagged monochromatic \kl\ beam is that of performing also lifetime measurements. The latter can be done by observing the time dependence of decays to a particular channel. Even better, the lifetime can be obtained by just counting all decays in a time, or space, interval.

Final results on \kl\ decays have been given by the KTeV collaboration \cite{ref:ktev}.
In their very complete study of semileptonic \kl\ decays, KTeV gives very accurate branching ratios and new precise values for the form factor parameters. The $K_{e3}$ branching fraction is found to be 0.4067\plm0.0011, strikingly different from the PDG fit value of 0.3878\plm0.0027. They have also clearly established the necessity of a curvature term in the vector form factor, quantitatively $\lambda''_+=\pt(3.20\pm0.69),-3,$. The linear term is however smaller,\footnote{Partial indications were given in \cite{ref:istramu}.} $\lambda'_+=\pt(20.64\pm1.75),-3,$. This results in a 1\% reduction of the phase space integral corresponding to an increase of 0.5\% for $|V_{us}|$. This reinforces the warning that comparison of results obtained with different parameter values are meaningless.

\section{The KLOE results}%%\makeatletter\global\@topnum\z@\makeatother
Another advantage of the monochromatic tagged \kl\ beam is the unambiguous closure of the kinematics. This however is not quite so important for just counting the number of different decay modes. There are two main cases to distinguish: 1) all neutral particles: \pio\po\ is \ab99.8\% of the total; 2) two charged particles final state, dominated by $\pi^\pm e^\mp\nu$, $\pi^\pm\mu^\mp\nu$ and \pic\po. The three contribution are quite well distinguished by the variable $\Delta_{\mu\pi}=E_{\rm miss}-|\vec p_{\rm miss}|$, computed using the muon and the pion masses for the charged particles and choosing the lowest value. Fig. 1 illustrates the channel separation as well as the agreement with Monte Carlo simulation which fully accounts for radiation in all channels.\\[3mm]
\figboxc fig1;6.6;\vglue2mm\noindent
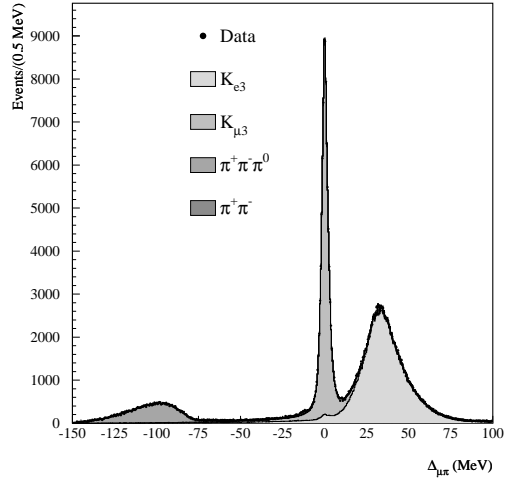
{Figure 1: \it Distribution of the variable $\Delta_{\pi\mu}=E_{\rm miss}-|\vec p_{\rm miss}|$, see text, for \kl\ decays to two charged particles. Only a few per cent of the data is plotted.}\vglue3mm
\noindent
The validity of this procedure is confirmed by using particle identification by time of flight and range-$\dif E/\dif x$ in the calorimeter. Fig. 2 shows an example of the $\Delta_{e\pi}$ distribution for events with an identified electron compared to MC prediction.\\[4mm]
\figboxc fig3;6.6;\vglue3mm\noindent
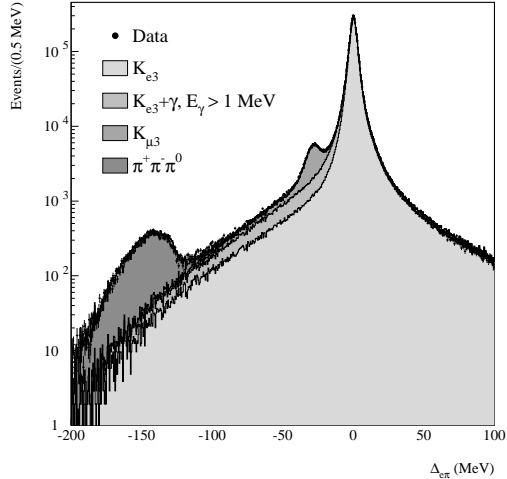
{Figure 2: \it Distribution of the variable $\Delta_{e\mu}$, see text, for \kl\ decays with identified electrons.}\\[3mm]
%\newpage\noindent
The final results for the branching ratios are:
{\makeatletter\abovedisplayskip 4pt plus1pt minus4pt\belowdisplayskip \abovedisplayskip
\abovedisplayshortskip \z@ plus1pt\belowdisplayshortskip 2.5pt plus.5pt minus1pt
\makeatother
\begin{eqnarray}
BR(\kl\to\eiii)&\kern-2.5mm=0.3985\pm 0.0035\\
BR(\kl\to\muiii)&\kern-2.5mm=0.2702\pm0.0025\\
BR(\kl\to\pio\po)&\kern-2.5mm=0.2010\pm0.0022\\ 
BR(\kl\to\pic\po)&\kern-2.5mm=0.1268\pm 0.0011
\end{eqnarray}}\hglue-0pt
The above results are obtained from the study of \ab13 million \kl\ decays. From the number of observed decays in the fiducial volume we obtain:
\begin{equation}
\tau(\kl)= 51.35 \pm 0.05_{\rm stat} \pm 0.26_{\rm syst}{\ \rm ns}
\end{equation}
An additional 40 million decays are used to obtain the various efficiencies. Details of the determination of BR(\kl\to\pio\po) are beyond the present discussion.

The entire data sample is used to measure the lifetime. Fig. 3 shows the time dependence of \kl\to\pio\po\ decays in the KLOE drift chamber and the region used in the lifetime fit. Regeneration in the beam pipe and inner chamber wall is visible.\\[1mm]
\figboxc tauklcol;7;\vglue2mm\noindent
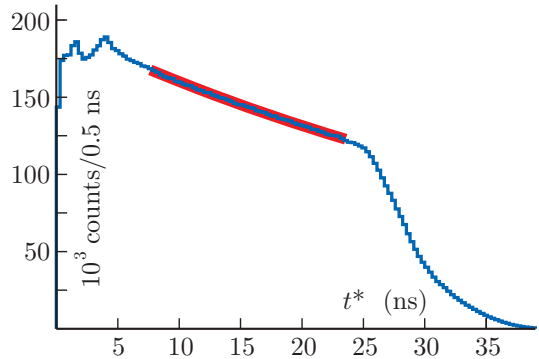
\cl{Figure 3: \it Distribution of the decay proper time for \kl\to\pio\po.}\vglue3mm
\noindent
We find:
\begin{equation}
\tau_{\kl}= 51.15 \pm 0.20_{\rm stat} \pm 0.40_{\rm syst}{\ \rm ns}
\end{equation}
These preliminary values are quite consistent with previous measurements \cite{ref:pdg} and together give a new improved lifetime: $\tau(\kl)=51.35 \pm 0.2$ ns.

\section{$|V_{us}|$}
Before getting to $|V_{us}|$, obtained from inverting eq. \ref{gamma1}, we wish to remark that experimental errors are getting smaller than 1\%. Specifically the lifetime contributes an error of 0.2\%, the BR values about 0.15\% and the form factor parameters some 0.3\%. Knowledge about the form factor normalization $f_+^{K^0}$ remains relatively poor. Leutwyler and Roos gave in 1984 \cite{ref:lr} the estimate $f_+^{K^0}$=0.961\plm0.008 which corresponds to a 0.83\% uncertainty. This result is confirmed by by lattice calculation \cite{ref:bec} and chiral perturbation theory \cite{ref:cir2,ref:bij} (up to unresolved order $p^6$ ambiguities \cite{ref:cms}) and we retain it in the following.
Table \ref{tab:vus} gives the final result for $|V_{us}|$, using the KTeV slopes and the KLOE lifetime for \kl. For $|V_{ud}|$ we use the value 0.9740\plm0.0005 of Czarnecki, Marciano and Sirlin \cite{ref:cms}.
\begin{table}[h]
\centering
\caption{$|V_{us}|$ from kaon semileptonic decay.}
\vskip 0.1 in
\begin{tabular}{|c|c|} \hline
  Source  & $|V_{us}|$ \\
\hline
\hline
 KTeV, \kl. $e3,\mu3$       & 0.2258\plm0.0013\plm0.0020   \\
 KLOE, \kl. $e3,\mu3 $ & 0.2248\plm0.0015\plm0.0020    \\
 KLOE, \ks. $e3$       & 0.2254\plm0.0017\plm0.0020   \\
 E865, $K^+$. $e3$            & 0.2288\plm0.0021\plm0.0020  \\
\hline
 Average              &  0.2259\plm0.0022  \\
\hline
$|V_{ud}|$ and unitarity &  0.2265\plm0.0022 \\
\hline
\end{tabular}
\label{tab:vus}
\end{table}
The first error for each entry is due to experimental uncertainties on branching ratios, lifetime and form factor parameters. The common error of 0.0020 is due to the error on $f_+(0)$. Charged kaons and neutral kaons corrections are from \cite{ref:cir1,ref:cir2,ref:lr}. The average is obtained by weighing according the first error. The $\chi^2$-value is 2.5, for 3 dof, mostly due to the $K^+$ result which is 1.5\sig\ above the average. The error is the quadrature of 0.0020, from $f_+(0)$, and 0.0008 from experiment. Unitarity is quite well satisfied: $\Delta=1-|V_{ud}|^2-|V_{us}|^2$=0.0003\plm0.0014, consistent with zero to 0.2\sig. The contributions to the error on $\Delta$ from the two terms are equal. Fig. 4 gives a different view of the $K^0$ results.\\[1mm]
\figboxc vusf0col;7;
\vglue4mm\noindent
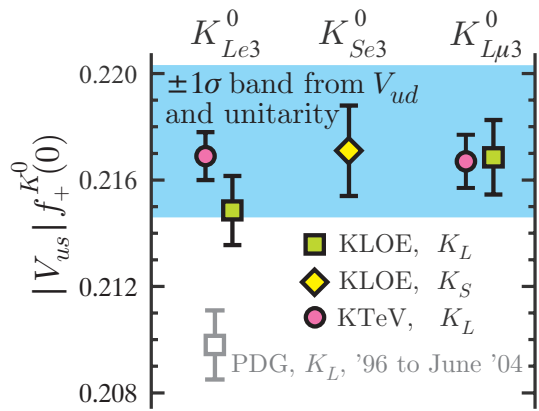
{Figure 3: \it The product $f^{K^0}_+(0)|V_{us}|$ for $K^0$ decays compared with unitarity expectation.}\vglue3mm
\noindent
This way of presenting the data has the advantage of showing more clearly their consistency, without discrepancies being hidden by the large error on $f^{K^0}_+(0)$. The error $f^{K^0}_+(0)$ is now reflected in the width of the ``unitarity'' band.

We do not include the result presented by Kleinknecht \cite{ref:klein} at HQ\&L2004 on June 2, because the $K_{e3}$ branching ratio is obtained using the old NA31 value for BR$(\kl\to\pio\po)$ \cite{ref:na31}. More recently \cite{ref:na48}, the NA48 presentation in Beijing at ICHEP04, using a judicious mixture of NA31 and KTeV values for the 3\po\ BR, gives, from the same data, the new preliminary value BR($K_{e3}$)=0.4010\plm0.0045.

{\makeatletter\def\section{\@startsection {section}{1}{\z@}{-2.5ex plus -.5ex minus
 -.2ex}{1.1ex plus .2ex}{\normalsize\bf}}\makeatother
\section{$|V_{us}|$ from $K\to\mu\nu$}
Marciano \cite{ref:mar} has recently pointed out how the new precise lattice \cite{ref:lat} results for the pseudoscalar meson decay constants, allow the determination of $|V_{us}|$ from the purely leptonic decays to muon-neutrino. In particular it is possible to relate the ratios $|V_{us}|/|V_{ud}|$ and $\Gamma(K\to\mu\nu)/\Gamma(\pi\to\mu\nu)$. In the notation of eq. \ref{gamma1} he gives:
\begin{equation}
\parbox{12cm}{\abovedisplayskip 4pt plus1pt minus4pt\belowdisplayskip \abovedisplayskip \abovedisplayshortskip 0pt plus1pt\belowdisplayshortskip 1pt plus.5pt minus1pt
$$\eqalign{
{\Gamma(K\to\mu\nu(\gam))\over\Gamma(\pi\to\mu\nu(\gam))}&={\left|{V_{us}\over V_{ud}}\right|^{\,2}}\kern-6pt\x{f_K^2\over f_\pi^2}\x I({\rm masses})\x(1+\delta_{\rm em})\cr
{|V_{us}|^2\over|V_{ud}|^2}&=0.05271\pm0.0151\cr
|V_{us}|&=0.2236\pm0.0036\cr}$$}
\label{eq:mu2}
\end{equation}
In eqs. \ref{eq:mu2}, $f_K/f_\pi=1.201\pm0.008\pm0.015$ comes from \cite{ref:lat}. The decay rates are from \cite{ref:pdg} and $|V_{ud}|$=0.9740\plm0.0005. From the previous discussion it is clear that the result for $|V_{us}|$ is in good agreement with unitarity and also with the results quoted before. It is interesting to note that this last $|V_{us}|$ value is low while the one obtained from $K^+\to\pi^0 e^+\nu$ is high. It is quite possible that the same cause is responsible for both effects. If the PDG $K_{\mu2}$ ``fit'' value is low, because of old inaccurate measurements and/or improper accounting for radiation, than the value of $|V_{us}|$ will of course be low. As already pointed out in sec. \ref{sec:new}, E865 has to use the PDG fit for the $K^+$ branching fraction in the $\pi2$, $\mu3$ and $\pi3$ modes. If BR($K_{\mu2}$) is low, then because of the fit $\pi2$+$\mu3$+$\pi3$ is correspondingly high, making the E865 final result also high by approximately $\sqrt{1.5}$ as much, when expressed in per cent.
KLOE has at present some preliminary indications that this might indeed be the case.
\section{Conclusion}
It is very gratifying to see the convergence of the new experimental results on kaon semileptonic decays. The CKM matrix appears now to be unitary and we can also use $\Gamma(K^\pm\to\mu^\pm\nu(\bar\nu)(\gam))$ and reach the same conclusion.
While the previously claimed violation of unitarity is certainly not there, it is however disappointing that $|V_{us}|$ is still only known to about 1\%. Agreement between different experiments is also still not as good as one could wish, but work is continuing. It is quite conceivable that in one or two years uncertainties due to lifetime, form factor shape parameters and the branching ratios will be brought to the 0.1\% level. We can only hope that a similar improvement will come about for $f_+(0)$. $\tau$ decays \cite{ref:tau} do not agree with the direct observation in kaon decays. This is similar to $\tau$ vs \epm\to\pic\ \cite{ref:klpp}.

\section{Acknowledgements}
I wish to acknowledge the help of Juliet Lee-Franzini through-out this entire work. I also acknowledge very helpful discussion with G. Isidori and W. Marciano. The KLOE young people of course did most of the work. Finally I thank Meenakshi Narain and Uli Heinz for their warm hospitality in Boston. I also thank the Alexander von Humboldt Foundation for support during part of this work.}%%\vglue-.2cm

\vglue-.1cm

\begin{thebibliography}{99}
\bibitem{ref:vub} This we learned in the early eighties at CESR.\vglue-.35cm

\vglue-.35cm
\end{thebibliography}
\end{document}